\documentclass[aps,prl,twocolumn,superscriptaddress]{revtex4-1}

\usepackage{amsfonts}
\usepackage{amsmath}
\usepackage{amssymb}
\usepackage{graphicx}
\usepackage{float}
\usepackage[dvipsnames]{xcolor}
\begin{document}

\title{Transmission of mobility via cooperative mechanisms in soft active matter.  }


\author{{\color{black}Victor Teboul}}
\email{victor.teboul@univ-angers.fr}
\affiliation{Laboratoire de Photonique d'Angers EA 4464, Universit\' e d'Angers, Physics Department,  2 Bd Lavoisier, 49045 Angers, France.\\
victor.teboul@univ-angers.fr}


\begin{abstract}

{\color{black}When supercooled, liquid's viscosity increases dramatically as the glass transition temperature is approached. While the physical origin of this behavior is still not understood, it is now well established that the addition of a few activated particles is able to reverse that increase in viscosity. 
Here we further raise the question of a limit in that fluidization process and of the differences between the fluidized liquid and its viscous counterpart.  
Results show that a few percent active molecules are enough to trigger a phase transition leading to diffusion coefficients typical of liquids while the medium retains cooperative properties of the viscous phase.
The similarity between cooperative properties of the active and non active molecules suggests
 that the mobility of active molecules is transmitted to inactive ones via the medium's cooperative mechanisms, a result in agreement with facilitation theories. 
 This result is then confirmed by the compared behavior of the distinct van hove correlation functions of most mobile active and non active molecules.
 Interestingly enough, 
 in our simulations the cooperative mechanisms are not induced or related to a decrease of the excitation concentration.
}

\end{abstract}

\maketitle
{\color{black}
\section{ Introduction}

Active matter\cite{active1} that is materials containing molecules able to move by themselves, like molecular motors\cite{motoro1,motoro2,motoro3,motoro4,motoro5,motoro6,motoro7,motoro8,motoro9,motoro10,motoro11,motoro12,motoro13,motoro14,motoro15,motoro16,motoro17,prefold,us1,md16} for example, retains a lot of attention due to its links with biology and out of equilibrium statistical mechanics.
More specifically, active matter could lead to a better understanding of the glass transition long standing problem\cite{gt0,gt1,gt2,anderson,fragile1,ms1,ms2,ms3,ms4,ms5}, due to its out of equilibrium nature {\color{black} and because active matter introduces a controlled cooperative mechanism inside the medium that may interfere \cite{PRE10} with the spontaneous one, an idea that we will follow here}.  A number of works, have been done in supercooled active matter\cite{active1,active2,active3,active4,active5,active6,active7,active8,active9,active10,active11,active12,active13,active14,active15,active16,active17,active18,active19,Szamel1,Szamel2,Szamel3,Szamel4,Szamel5,Szamel6,Szamel7,PRE10}. It was found that adding active particles to a viscous medium permits to decrease its viscosity without significantly increase its temperature, an effect called fluidization\cite{flu1,flu2,flu3,flu4,flu5,flu6,md16,cage,prefold,rate,us1,md16,rate,c3,c4,pccp,carry,PRE10}.

Is it possible to liquefy entirely a solid depending on the concentration of active particles ? and what is the physical origin of that process ?
To answer these questions we increase the concentration of active molecules while the temperature of the medium is kept constant.
We find in this work that above a concentration threshold the medium's viscosity actually drops.
We find that the mobility of active molecules is transmitted to inactive ones via the medium's cooperative mechanisms\cite{md16,PRE10,dh0,dh1,dh2,dh3}, a result  in agreement with facilitation theories\cite{facile,facile1,facile2,facile3,facile4,facile5,facile6}.
The transmission then continues from inactive ones to other inactive molecules via the cooperative mechanisms until the transition is reached if the threshold of active molecules concentration is attained.
We find that cooperative mechanisms do not decrease when the concentration of active molecules, then the concentration of excitations increases, as long as the temperature is maintained constant.
This result shows that spontaneous cooperative mechanisms in supercooled liquids are not induced or directly related to the decrease of the excitation concentration  when the temperature decreases.
The spontaneous cooperative motions are more probably induced by the decrease of the kinetic energy (temperature) that induces a preponderance of the potential energy, therefore an organization of the medium. 

\section{Calculation}

Due to the importance of the cage effect in supercooled liquids the instantaneous velocity is not a relevant physical parameter, as the molecules bouncing in the cages created by their neighbors do change frequently their velocity direction before the physically relevant cage escaping process. The physics of the system is therefore better captured by coarse graining in time the velocity using a time difference corresponding to the average cage escaping time. This parameter is called the mobility.
More precisely we define the mobility $\boldsymbol \mu_{i}(t)$ of a molecule $i$ as 
\begin{equation}
\displaystyle {\boldsymbol \mu_{i}(t)={\bf r}_{i}(t+\tau_{m})-{\bf r}_{i}(t)}    \label{e0}
\end{equation}
where $\tau_{m}$ is the mobility characteristic time.
\vskip0.5cm

A common approach to simulate active media involves applying a propulsive force to molecules, typically aligned with their velocities. Sometimes, additional interactions that encourage alignment of molecular displacements are introduced to simulate complex behaviors akin to biological movements.
In this study, we explore the effects of forces that more closely resemble the physics of supercooled liquids. Consequently, instead of aligning with velocities, our propulsive force will track the mobility of molecules, and correlations between velocities will be replaced by correlations in molecular mobilities. In essence, by activating our medium with relevant parameters, we anticipate an increase or decrease in the thermal cooperativity of the medium as it approaches the glass transition. This, in turn, should lead to modifications in the transition and physical properties of the medium, particularly fluidization, which we aim to investigate.
To achieve this, we employ out-of-equilibrium molecular dynamics simulations\cite{md1,md2,md2b,Aurelien} for results that are both practical to interpret and easy to compare with previous studies on molecular motors \cite{md16,rate,c3,c4,pccp,carry}. Specifically, that method will make easier to compare our findings to experimental and theoretical observations of fluidization phenomena\cite{md16,flu1,flu2,flu3,flu4,flu5,cage,flu6}  when molecular motors are dispersed within soft materials.
Our medium is a minimal model liquid\cite{ariane} chosen to hinder crystallization and accelerate the simulations. It is constituted of dumb bell diatomic molecules, with rigidly bonded atoms (indexed as $i=1, 2$) with a fixed interatomic distance of $l = 1.73$ Å. The two atoms are defined with the same mass $m_{0}=20g/N_{A}$.    
These atoms interact with atoms from other molecules through the following Lennard-Jones potentials:
\begin{equation}
V_{ij}=4\epsilon_{ij}((\sigma_{ij}/r)^{12} -(\sigma_{ij}/r)^{6})   \label{e1}
\end{equation}
with the parameters\cite{ariane,mix1,mix2}: $\epsilon_{11}= \epsilon_{12}=0.25 KJ/mol$, $\epsilon_{22}= 0.2 KJ/mol$,  $\sigma_{11}= \sigma_{12}=3.45$\AA, $\sigma_{22}=3.28$\AA.
The length of the molecule is therefore $l_{m}=5.09$\AA\ and its width $L_{m}=3.37$\AA.
Our cubic simulation boxes contains $1000$ molecules and are $32.49$ \AA\ larges, or $2000$ molecules and are $40.9$ \AA\ larges.

The system is maintained out of equilibrium by the presence of active molecules releasing energy into it.
A  thermostat\cite{berendsen}  removes the energy dissipated into the system avoiding a  drift in energy. 
In our calculations,  a few percents of the medium molecules are activated (pushed) in the mobility direction of the most mobile of their neighbors during $10 ps$ taken at random but continuously in a $40 ps$ time lapse. Therefore, at any time  a few percents are accelerated, while the other molecules do not experience external forces but only the intermolecular interactions. 
Within this work we will call active, the molecules that are periodically activated, and nonactive the other molecules. 
Activated molecules are periodically subject to a force ${\bf {f}}_{i}^{a}$ of constant intensity $f_{0}$, acting during $10 ps$ of a time period $T=40ps$. The forces ${\bf{f}}_{i}^{a}$  begin with a different random time origin for each active molecule, following the law:
\begin{equation}
 {\bf {f}}_{i}^{a}({\tau_{m}},t) = {{ f_{0} {\theta}_{i,T,{\Delta T}}(t)  {\bf{u}}_ {i,{\mu}_{max}}^{neighbor}}(t,{\tau}_{m})}            \label{e1}
\end{equation}
Here ${\theta}_{i,T,{\Delta T}}(t)$ is a periodic heaviside function, equal to $1$ during $\Delta T=10ps$ and zero otherwise with a period $T=40ps$.
$f_{0}=3.01$ $10^{-14} N$ is the constant intensity of the force when activated.  
\begin{equation}
{ {\bf{u}}_ {i,{\mu}_{max}}^{neighbor}(t,{\tau}_{m})=\boldsymbol \mu_{j}(t,{\tau}_{m})/\lvert \mu_{j}(t,{\tau}_{m})\rvert}        \label{e1b}
\end{equation}
 is the unit vector of the mobility of the most mobile neighbor $j$ of molecule $i$.

In this study, we investigate the alteration of key features in supercooled liquids, such as dynamic heterogeneity, diffusion properties, and the $\alpha$ relaxation time, which is associated with the viscosity of the medium. We will now define the statistical functions utilized for this purpose. One function of significant relevance in glass-transition phenomena is the intermediate scattering function, denoted as $F_{S}(Q,t)$, which portrays the autocorrelation of density fluctuations at the wave vector $Q$. This function provides insights into the structural relaxation of the material. We define $F_{S}(Q,t)$ through the following relation:

\begin{equation}
\displaystyle{F_{S}(Q,t)={1\over N N_{t_{0}}} Re( \sum_{i,t_{0}} e^{i{\bf Q.(r_{i}(t+t_{0})-r_{i}(t_{0}))}}  )          }\label{e1}
\end{equation}
For physical reasons, Q is chosen as the wave vector (here $Q_{0}=2.25$\AA$^{-1}$) corresponding to the maximum of the structure factor $S(Q)$.
$F_{S}(Q_{0},t)$ then allows us to calculate the $\alpha$ relaxation time $\tau_{\alpha}$  of the medium from the equation: 
\begin{equation}
\displaystyle{F_{S}(Q_{0},\tau_{\alpha})=e^{-1}}       \label{e10}     
\end{equation}
Finally, the diffusion coefficient $D$ is obtained from the long time limit of the mean square displacement $<r^{2}(t)>$:
\begin{equation}
\displaystyle{<r^{2}(t)>=   {1\over N N_{t_{0}}}  \sum_{i,t_{0}}  ({\bf r}_{i}(t+t_{0})-{\bf r}_{i}(t_{0}))^{2}                                }       \label{e11}     
\end{equation}
and
\begin{equation}
\displaystyle{\lim_{t \to \infty}  <r^{2}(t)>=  2 d D t                               }       \label{e12}     
\end{equation}

In this work we will chose $\tau_{m}=200 ps$ unless otherwise specified and the dimension $d=3$.


\section{Results and discussion}

\begin{figure}
\centering
\includegraphics[height=7.5 cm]{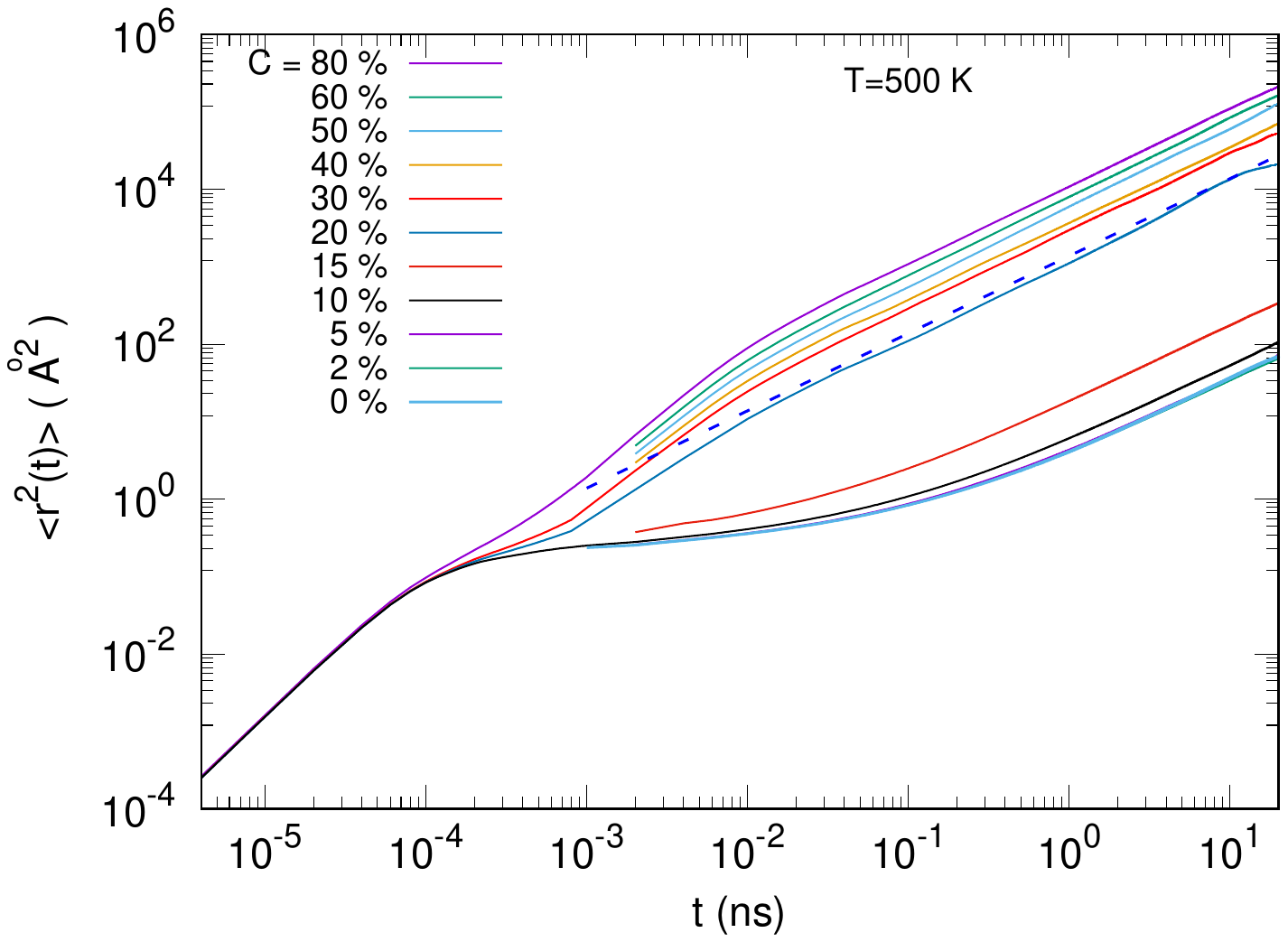}

\caption{(color online)  Mean square displacement $<r^{2}(t)>$ time evolution for various concentration of active molecules. In blue dashed line, the experimental diffusion of liquid water at ambient pressure and temperature is displayed for comparison.}
\label{f1}
\end{figure}

When increasing the concentration $C$ of active molecules, Figure \ref{f1} shows that, below a phase transition that takes place around the critical concentration $C_{c}=20\%$ for $\tau_{m}=200 ps$, the mean square displacement (MSD, $<r^{2}(t)>$) follows the typical behavior of supercooled liquids. Below $C_{c}$, for curves at the bottom of the Figure, the MSD displays a ballistic behavior at short time scales ($t<10^{-4} ns$) followed by a plateau at intermediate time scales (from $10^{-4}$ns to approximately $10^{-1}$ ns) and a diffusive behavior at large time scales ($t>1ns$).
In this range of concentration of active molecules we observe a regular increase of the MSD with the concentration $C$, while the overall MSD behavior remains unchanged.  However, this increase in MSD remains limited (it does not exceed a factor  4).
 Then for $C=C_{c}=20\%$ and beyond, we observe a significant change in the behavior of the MSD as well as a strong increase (by a factor greater than $10$) in the diffusive part. Although the ballistic behavior at short times and the diffusive behavior at long times persist, the plateau corresponding to the part of the trajectory where the particle is confined in the cage created by its neighbors is greatly modified. We observe a curvature inversion in the time range of the former plateau, resulting in a diffusion regime that occurs more rapidly. 
We interpret this behavior as resulting from molecules escaping faster from their cages due to the presence of nearby active molecules.
{\color{black} More generally, the disappearance of the plateau is a signature of the disappearance of the cage effect, resulting in a liquid like behavior. While there is no global structural signature of that disappearance, it can however be connected to the aggregation of  active molecules resulting in the medium's fluidization.}
Above the transition, the medium is more diffusive than the non activated medium at a temperature twice larger {\color{black} or than liquid water at room temperature (dashed lines)}.

\vskip 0.5 cm

\begin{figure}
\centering
\includegraphics[height=7.5 cm]{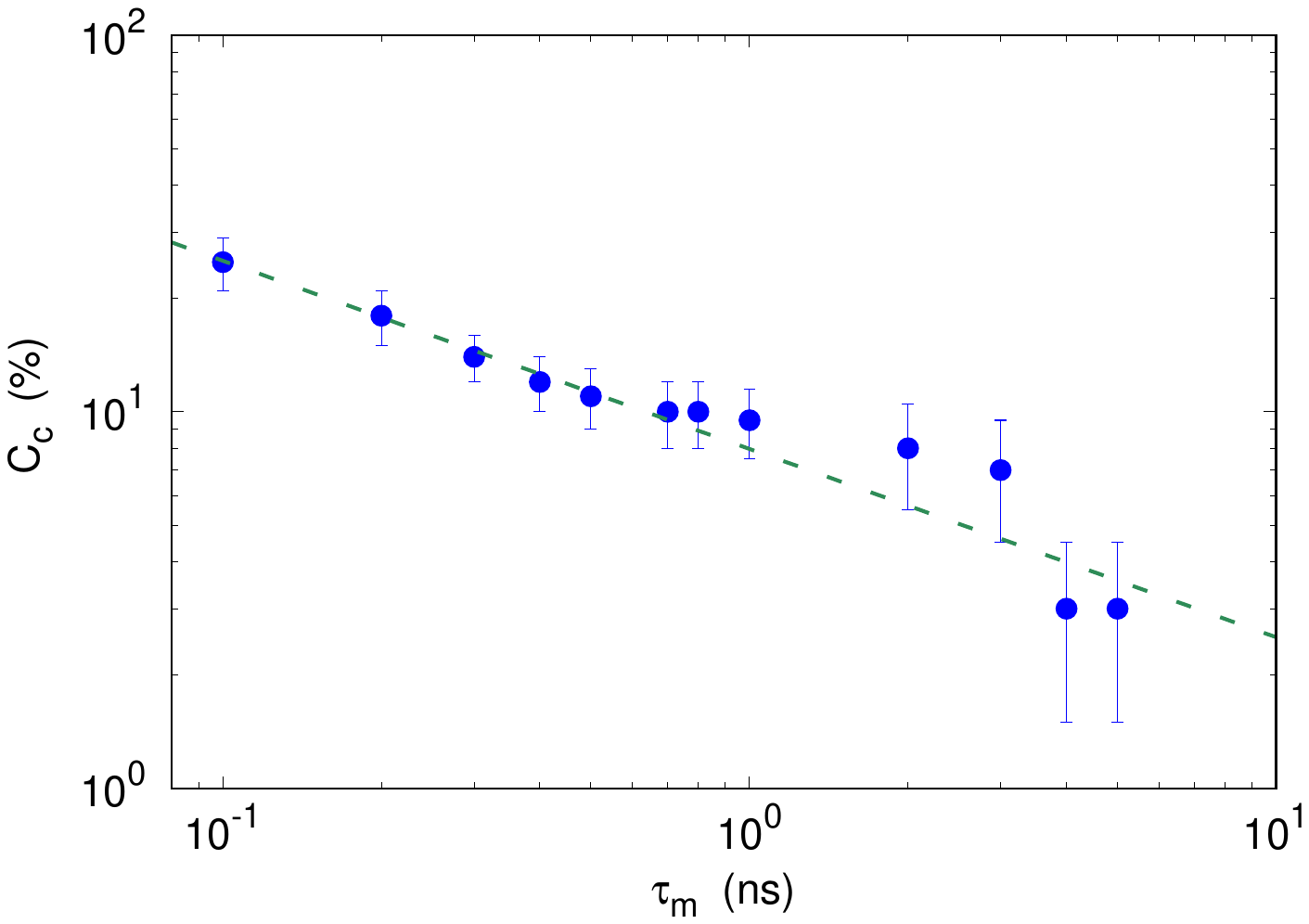}
\caption{(color online)  Active particules concentration threshold $C_{c}$ for the transition versus the characteristic time  $\tau_{m}$ used for the mobility. We observe a decrease of the minimum number of mobile molecules necessary to trigger the transition as $\tau_{m}$ increases. The line is a fit evolving as $C_{c}=C_{0}.\tau_{m}^{-0.5}$.}
\label{f0}
\end{figure}

Figure \ref{f0} shows the dependence of the critical concentration of active particules on the mobility characteristic time $\tau_{m}$. The Figure shows that as the mobility characteristic time $\tau_{m}$ increases the system needs less active particules to undergo its transition to a liquid.
At the right end side of the curve, the critical concentration reaches $3 \%$. As only $1/4$ of the active molecules are activated at a given time $t$, that means that less than $1\%$ of activated molecules are enough to transform the viscous medium into a liquid. We interpret that result as a larger coupling between activated particules and the medium when the mobility time approach the characteristic time of the medium's spontaneous cooperative mechanism $t^{*}(T)$ at the temperature $T$ of study.

\vskip 0.5 cm
 \begin{figure}
\centering
\includegraphics[height=7.5 cm]{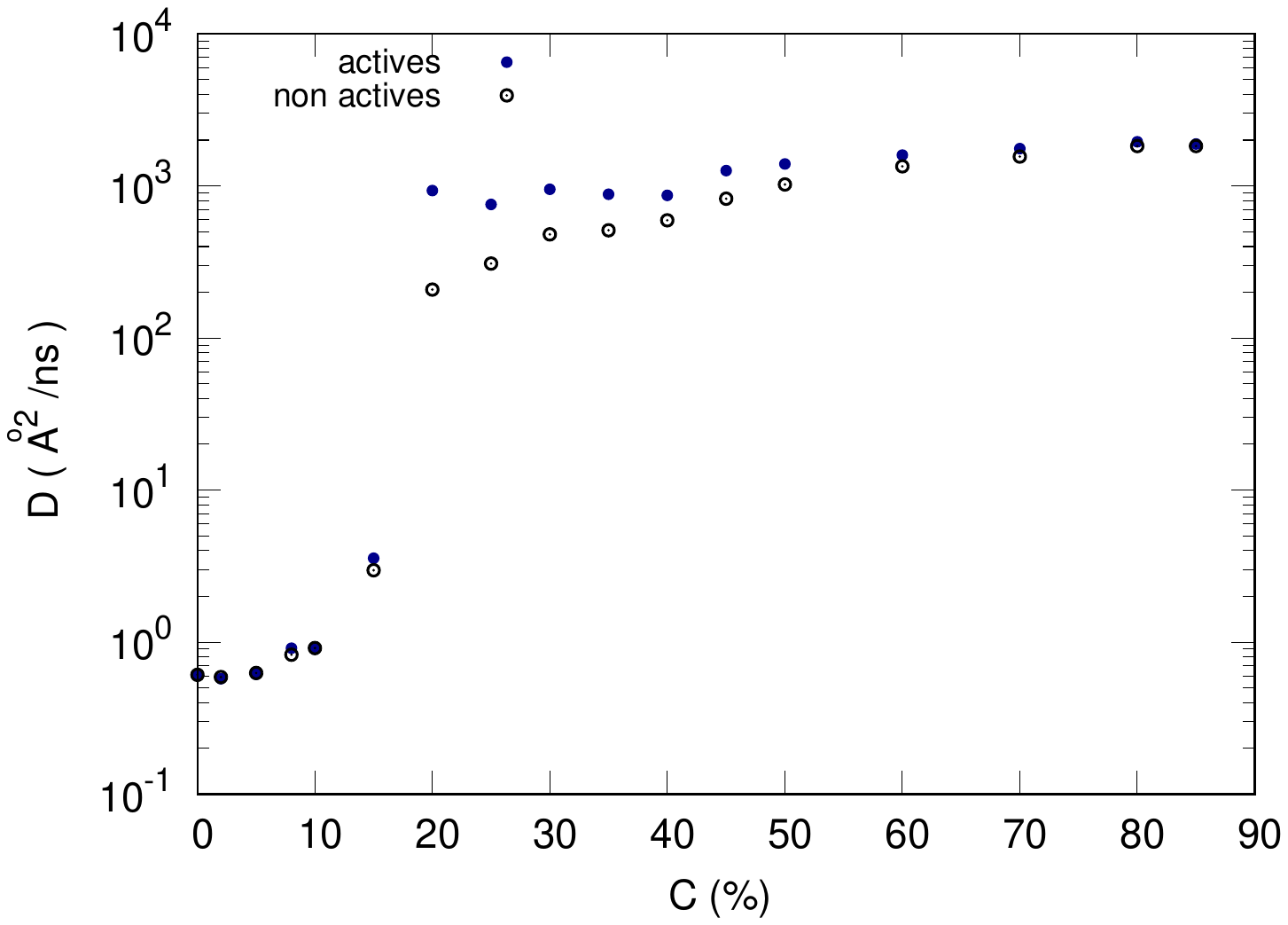}
\caption{(color online) Diffusion coefficient $D$ for active and non active molecules versus the concentration C of active molecules. We observe a transition from a soft solid to a liquid diffusion coefficient when C reaches 20 \%.}
\label{f2}
\end{figure}

The diffusion coefficient evolution with the concentration of active molecules $C$, displays in Figure \ref{f2} a rapid increase when $C$ reaches $C_{c}$, corresponding to the dynamic phase transition observed recently \cite{PRE10} in a similar system.
The transition is sharper for active molecules than non actives (see Figure \ref{f2}).  That result suggests that active molecules drive the dynamics of the non active molecules. In other words, our results suggest that the mobility propagates from active to non active molecules.
However, the diffusion jump occurs at the same threshold concentration of 20\%, which demonstrates that the active components alone cannot undergo transition, but require the inactive ones. This leads us to the idea that the active components propagate their activity to the inactive ones (via the cooperativity of the medium, a result we will see further on) until there are (when the concentration of active components increases) enough mobile molecules in the medium for the transition to occur. The fact that the diffusion coefficient changes little afterwards (i.e. above $C_{c}$) indicates that all molecules become mobile when the threshold is reached.

\begin{figure}
\centering
\includegraphics[height=7.5 cm]{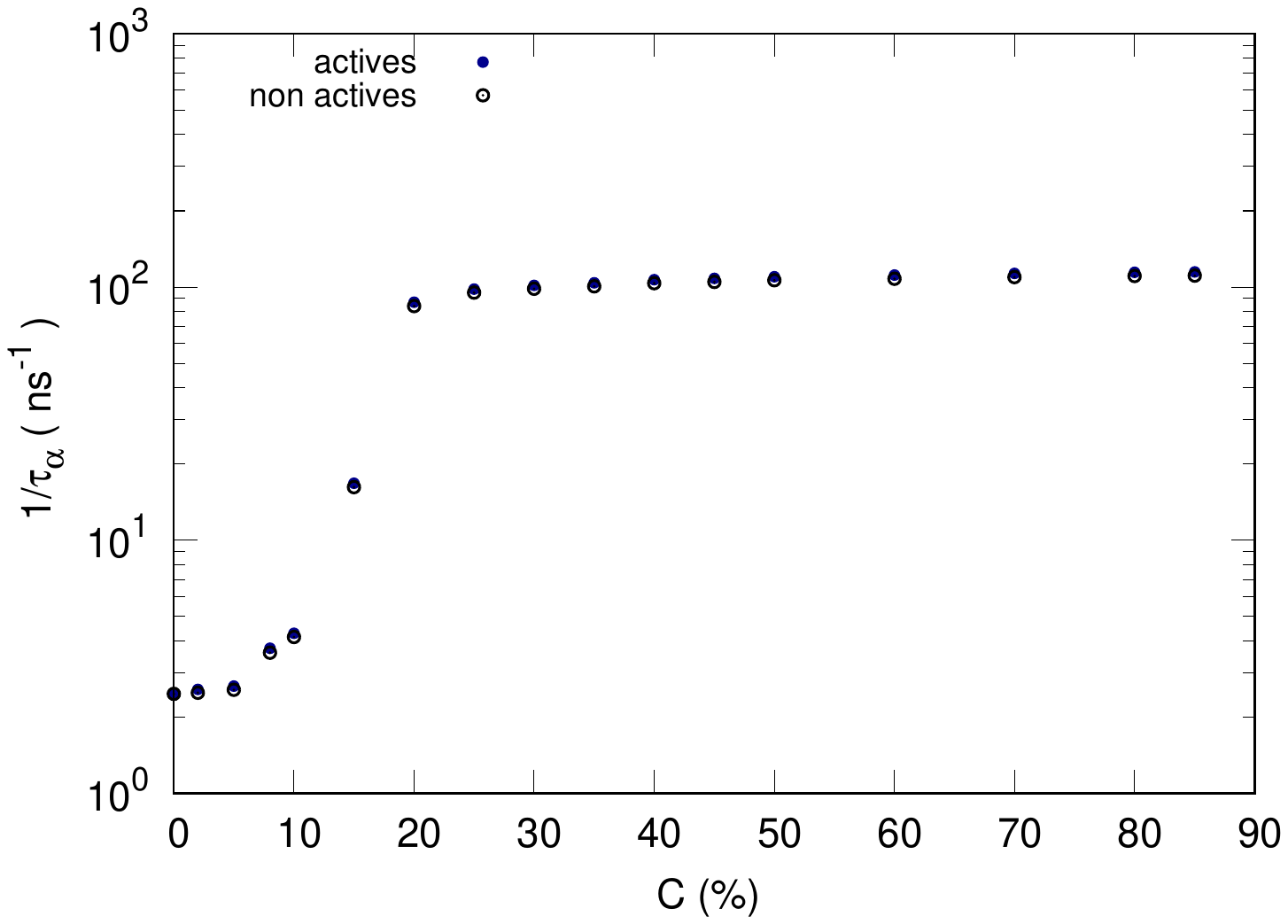}

\caption{(color online)  Inverse of the $\alpha$ relaxation time versus the concentration C of active molecules. }
\label{f3}
\end{figure}

This result suggests that dynamic heterogeneities (cooperative movements of molecules) are the origin of the transition. In other words, the intrinsic cooperativity of the medium transmits the mobility of active molecules to inactive molecules. We will now test this hypothesis.
The $\alpha$ relaxation time $\tau_{\alpha}$ is proportional to the viscosity of the medium, i.e., the local behavior of the medium, while diffusion, is linked to the medium's global behavior. Due to that difference, the diffusion coefficient and the $\alpha$  relaxation time are affected differently by heterogeneities.
Figure \ref{f3} shows the same transition for $\tau_{\alpha}$ than for the diffusion coefficient in Figure \ref{f2}, but this time both active and inactive components follow exactly the same behavior. This difference suggests that the more significant increase in the diffusion of active components is due to heterogeneous behaviors.
It has indeed been demonstrated that due to the local nature of viscosity, significant movements of a small number of molecules lead to more pronounced effects on diffusion than on viscosity, making diffusion more sensitive to heterogeneous behaviors. Viscosity tends to be governed by less mobile molecules on average, while diffusion is more affected by highly mobile molecules.
{\color{black} As $\tau_{\alpha}$ is related to the viscosity, we interpret the transition in figure \ref{f3} as a fluidization transition.}

\begin{figure}
\centering
\includegraphics[height=7.5 cm]{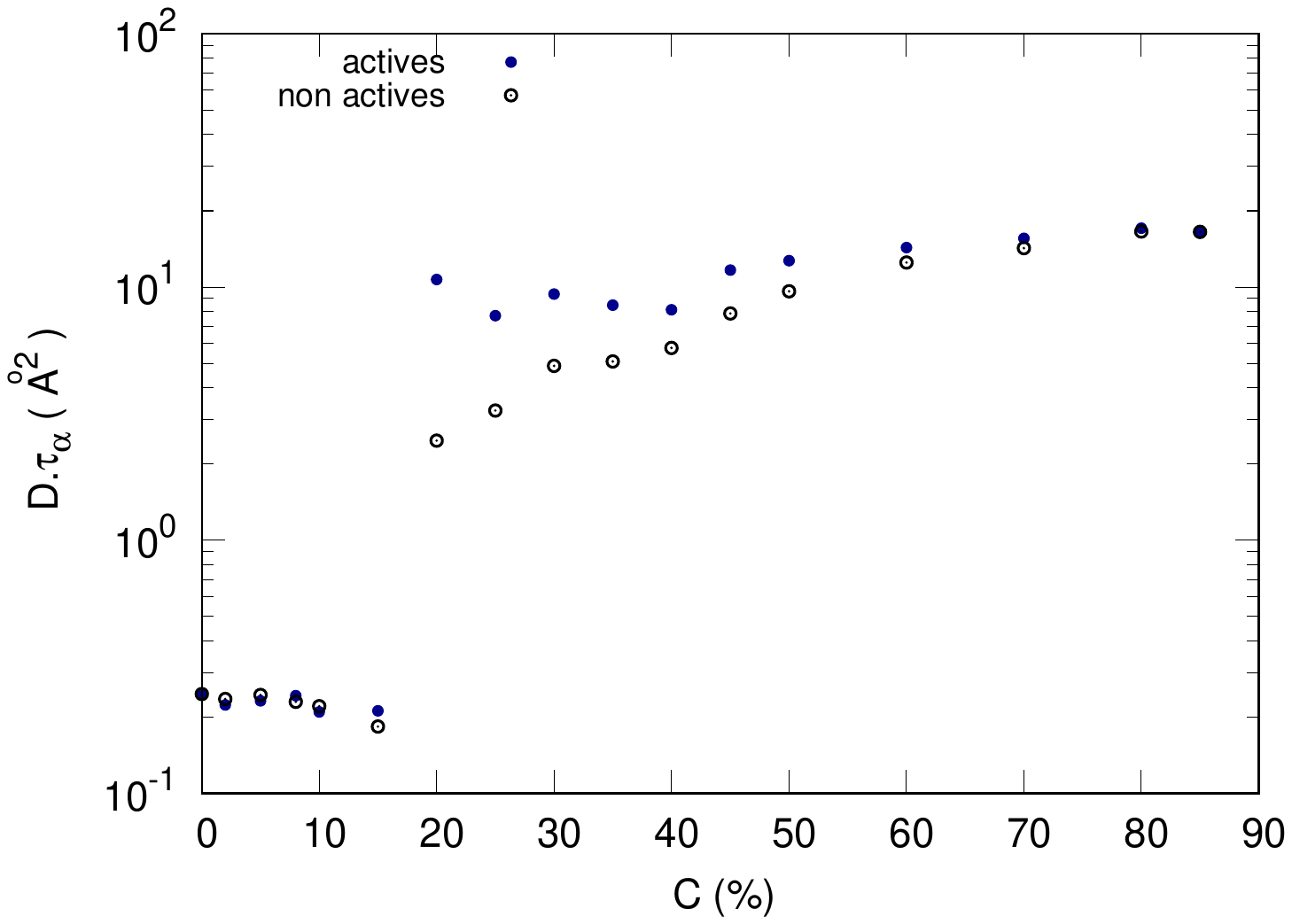}

\caption{(color online)  Breaking of Stokes Einstein law, measured from the product of the diffusion coefficient and of the $\alpha$ relaxation time: $D.{\tau_{\alpha}}$. }
\label{f4}
\end{figure}

The heterogeneous nature of the dynamics can be tested through deviations from the Stokes-Einstein law, which relates diffusion to viscosity. The law implies that in a normal liquid, the product $D.\tau_{\alpha}/T$ should remain constant. Here, we have omitted the temperature because $T$ is constant in our simulations. Deviations from the law appear in supercooled liquids, which are explained by the emergence of heterogeneous cooperative dynamics.
We observe in Figure \ref{f4} that at low concentrations of active molecules, the dynamics remain heterogeneous and cooperative. Just before the transition, we observe a slight but significant decrease in cooperativity, followed by a significant increase in cooperativity at the transition point. This cooperativity then evolves relatively little for higher concentrations of active molecules. This behavior and transition are more pronounced for active molecules than for inactive ones.
For inactive molecules, cooperativity continues to evolve above the transition point as the concentration C increases, tending towards the value of active molecules. The behavior observed in the figure suggests that active molecules drive the dynamics of inactive ones, but the transition cannot occur without the contribution of inactive molecules. The phenomenon of slight decrease in cooperative motions just before the transition remains unexplained.
This result suggests the existence of a counter-reaction mechanism. {\color{black} A tentative explanation is that the local increase in diffusion washes out the long range cooperative motions.}

\begin{figure}
\centering
\includegraphics[height=7.5 cm]{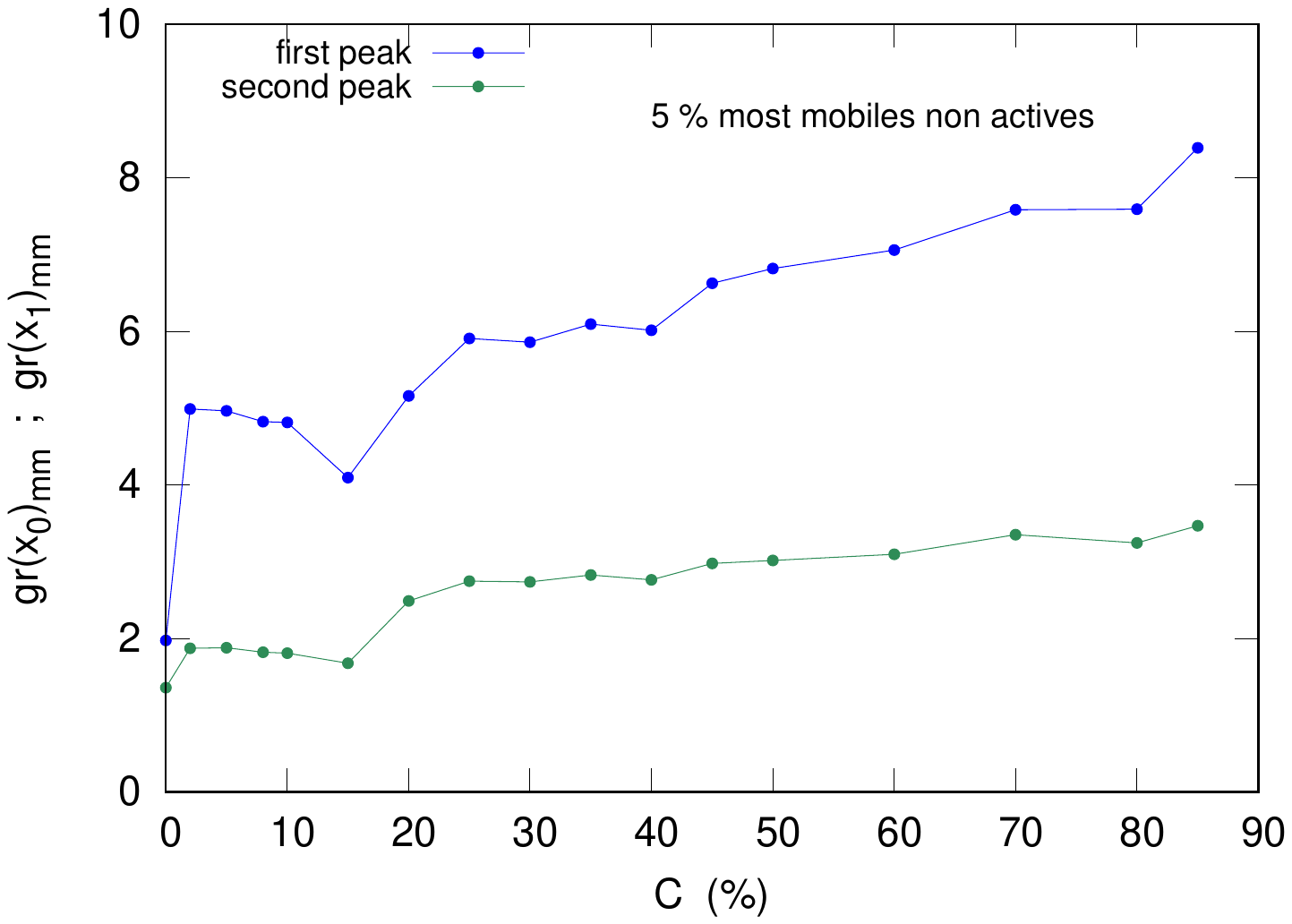}

\caption{(color online)  Maximum values of the first and second peaks of the radial distribution function of most mobile non active molecules. }
\label{f5}
\end{figure}
\begin{figure}
\centering
\includegraphics[height=7.5 cm]{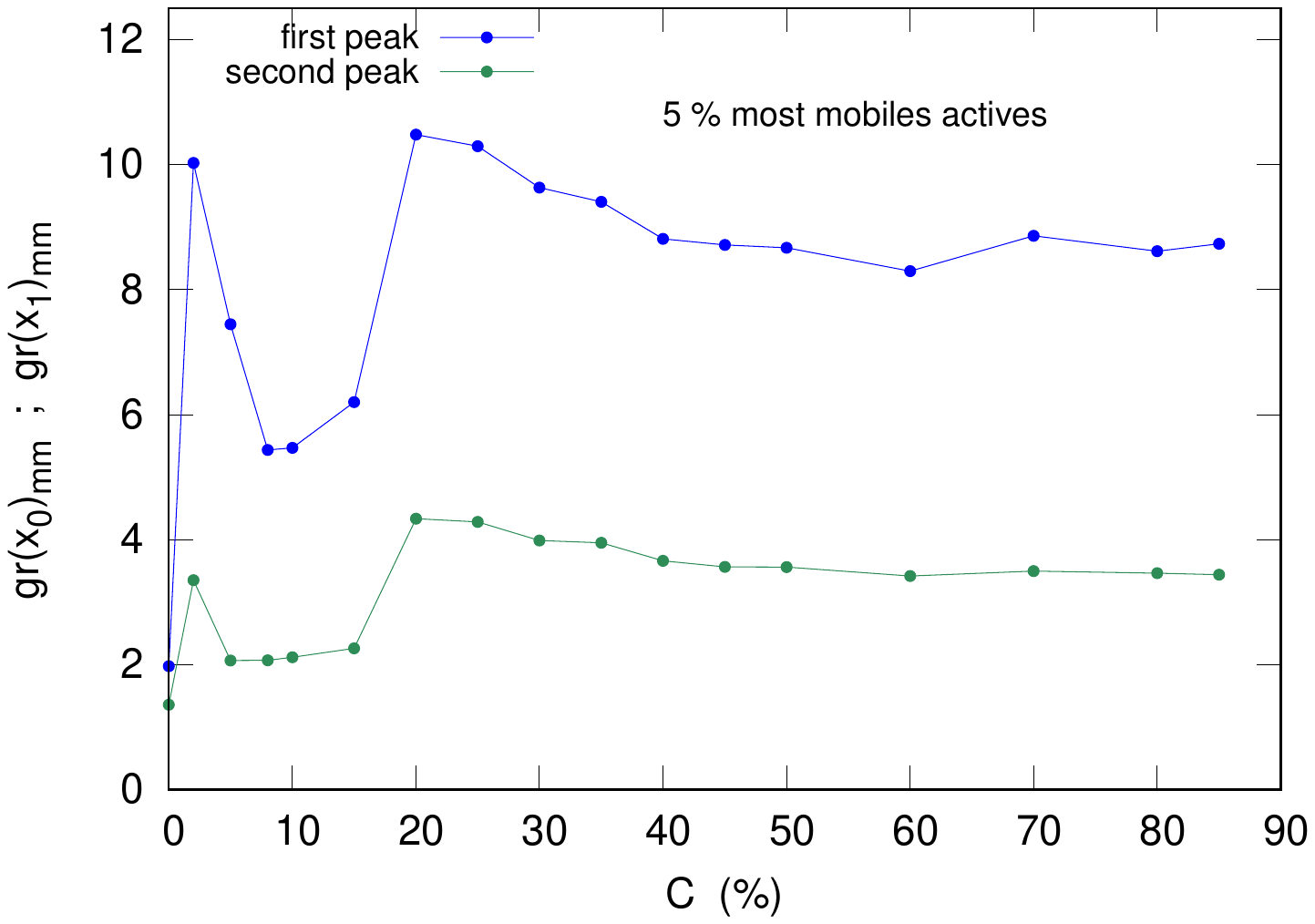}

\caption{(color online)  Maximum values of the first and second peaks of the radial distribution function of most mobile active molecules. }
\label{f5b}
\end{figure}

To confirm these results, we will now study other characteristic quantities of cooperativity in supercooled media. One of these characteristics is the aggregation of mobile molecules, which can be observed in the radial distribution function restricted to mobile molecules. The first and second peaks of this function increase due to the dynamic aggregation of mobile molecules which is a typical feature of cooperative mechanisms in supercooled liquids. The first point ($C=0$) corresponds to the average radial distribution function without active molecules. This function does not change as the concentration of active molecules increases and serves as a reference for cooperativity.
In Figure \ref{f5}, we observe a relatively constant aggregation followed by a decrease just before the transition, then a continuous increase in the maximum of the first peak, with the second peak remaining approximately constant after the transition. Thus, we observe a behavior similar to that observed for the deviation from the Stokes-Einstein law, confirming that the deviation from the SE law arises from cooperativity, here measured via the aggregation of mobile molecules. The sharp decrease of cooperative motions just before the transition is also confirmed here. If the results we present in this article concern, for the sake of clarity, simulations using a particular time $\tau_{\mu}= 200 ps$, we found that sharp decrease in our calculations for multiple values of $\tau_{\mu}$. More importantly, that decrease appears to be larger for active than inactive particles as observed in Figure \ref{f5b}, suggesting that this effect originates in active molecules and propagates to inactive ones.
{\color{black} The observed aggregation of mobile molecules, here upon activation of the medium, is considered as the main signature of dynamic heterogeneity (cooperative motions). Due to the facilitation mechanism, molecules motion is facilitated near already mobile molecules and the aggregation thus induces a fluidization of the medium.}

\begin{figure}
\centering
\includegraphics[height=7.5 cm]{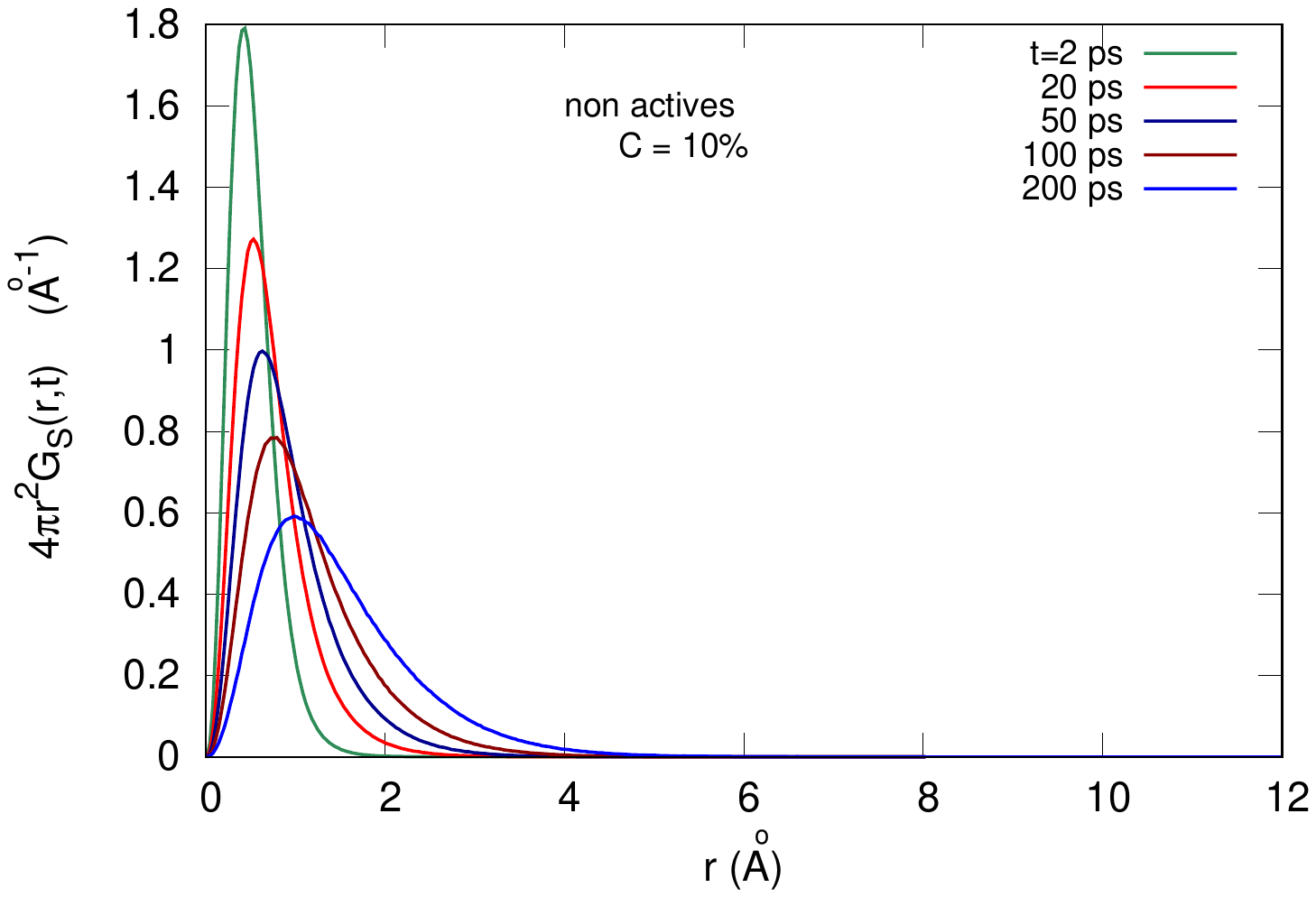}

\caption{(color online)  Self part of the Van Hove correlation function below the transition. A small tail can be observed for $t=200 ps$.}
\label{f6}
\end{figure}

To better understand the dynamics of the medium and confirm cooperative effects, we will now study the self and distinct Van Hove functions for inactive molecules. Let's start with the self Van Hove functions (Figures \ref{f6} and \ref{f7}). These functions describe the probability distribution of finding a molecule at a distance r from its initial position after a time lapse $t$. They therefore describe the distributions of displacements. Below the critical concentration $C_{c}$, we observe in Figure \ref{f6}, for time intervals ranging from about 20 to 200 ps, the appearance of a small tail in the correlation function, indicating the presence of a small number of molecules moving with greater mobility. This feature is characteristic of cooperative movements in supercooled media, where a few number of molecules moving cooperatively exhibate larger displacements.

 \begin{figure}
\centering
\includegraphics[height=7. cm]{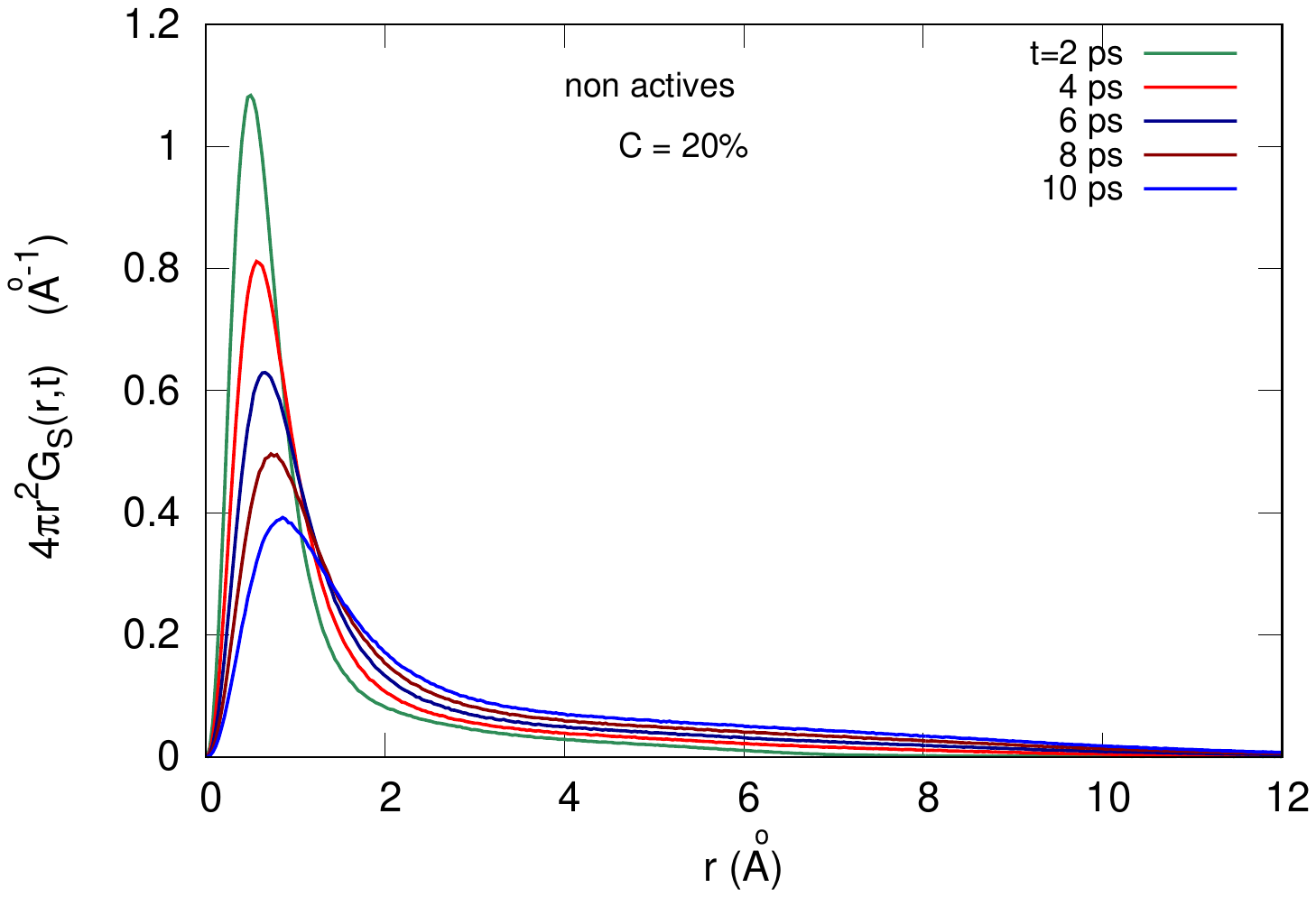}

\caption{(color online)  Self part of the Van Hove correlation function above the transition. The tails corresponding to cooperative displacements are much larger than below the transition.}
\label{f7}
\end{figure}

If we now compare these results to what happens at the transition or above, we see in Figure \ref{f7} that above the transition the tail appears on shorter timescales, is much more pronounced, and extends over much greater distances. The effect of the transition is therefore to significantly increase the displacement of the most mobile molecules. This result again shows an increase in cooperative movements of molecules at the transition, as the Van Hove tail is a fingerprint of cooperative motions in supercooled liquids. 
This result is quite intriguing as above the transition the diffusion coefficient also increases sharply, and the medium is liquid-like. Cooperative motions are therefore not related to viscous dynamics in our case, but purely induced by the activation mechanism.

{\color{black}
 \begin{figure}
\centering
\includegraphics[height=7. cm]{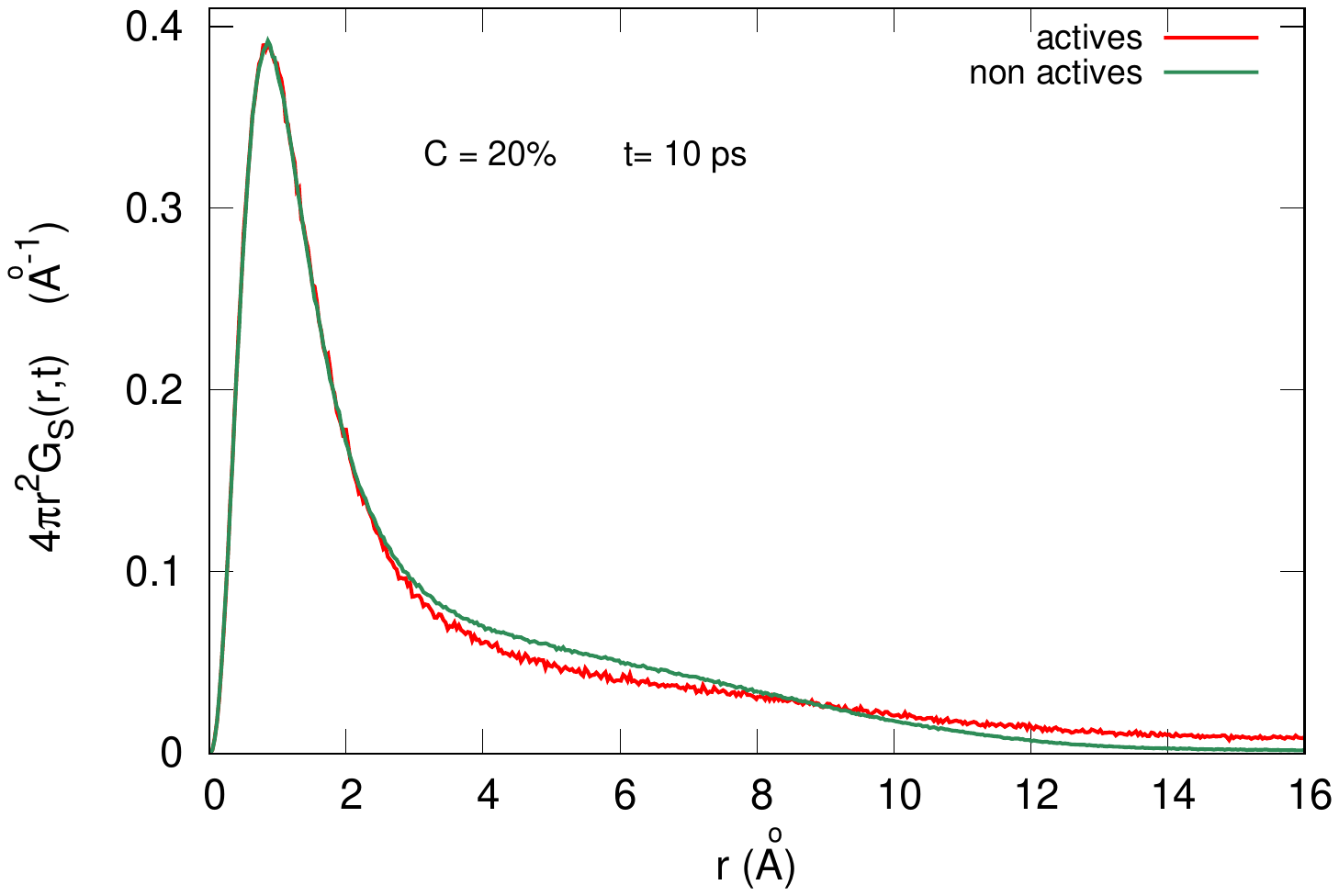}

\caption{(color online)  Comparison between active and non-active molecules self part of the Van Hove correlation function above the transition. }
\label{f7c}
\end{figure}

In Figure \ref{f7c} we compare the self Van Hove correlation functions for active and non active molecules above the transition.
We observe on the Figure the exact same curve for active and non active molecules for $r<3 $ \AA\ while the curves differ at larger travelled distances.
At large distances (i.e. for large mobilities) the non active curve tail is shorter, showing larger cooperative motions for the active molecules.
The fact that the non active tail appears when the number of active molecules is large enough (Figures \ref{f6} and  \ref{f7}) and the comparison of the active and non active Van Hove tails in Figure \ref{f7c}, suggests again that active molecules propagate their cooperative movements to inactive molecules. 
To confirm that result we will now study the distinct part of the Van Hove correlation functions for the most mobile molecules in Figure \ref{f8a}. 
}

{\color{black}
Let's focus now on the distinct part of the Van Hove correlation function $G_{d}(r,t)$. This function describes the probability distribution of finding a molecule at a distance r from the initial position of the other molecules after a time lapse $t$.
The distinct part of the Van Hove correlation function $G_{d}(r,t)$ restricted to the $6 \%$ most mobile molecules,  in Figure \ref{f8a} show an increase in the function around $r=0$ over a characteristic time $\Delta t$. 
That feature is a characteristic of dynamic heterogeneities in supercooled viscous media and shows the presence of chain-like movements, a molecule replacing another one on the characteristic time $\Delta t$.

Below the transition for $C=10\%$ (Figure \ref{f8a}(a)) the chain-like motions are similar between active and between non active molecules  while it is larger between active and non-active molecules, showing the importance of the propagation of cooperative motions from the active to the non active molecules.

When we increase $C$ to $15 \%$ (Figure \ref{f8a}(b)) just below the transition or above the transition to $20 \%$ (Figure \ref{f8a}(c)), the cooperative characteristic time $\Delta t^{*}$ for a molecule to be replaced by one of its neighbors drops fro $\Delta t^{*}= 420 ps$ to $\Delta t^{*}=110 ps$  ($C=15 \%$) and eventually $\Delta t{*}=2 ps$ above the transition ($C=20 \%$).
This sharp decrease of $\Delta t^{*}$ leads to a decrease of the $\alpha $ relaxation time $\tau_{\alpha}$ and therefore of the medium's viscosity.
Figures \ref{f8a} (b) and (c) also show that around and above the transition the active molecules replace mostly other active molecules in their cooperative motions while the remaining active molecules being replaced by non active ones have the same proportions than non active being replaced by other non actives. 
These results like the results observed below the transition emphasize the active molecules origin of the whole medium's cooperativity. 

}

\begin{figure}
\centering
\includegraphics[height=7. cm]{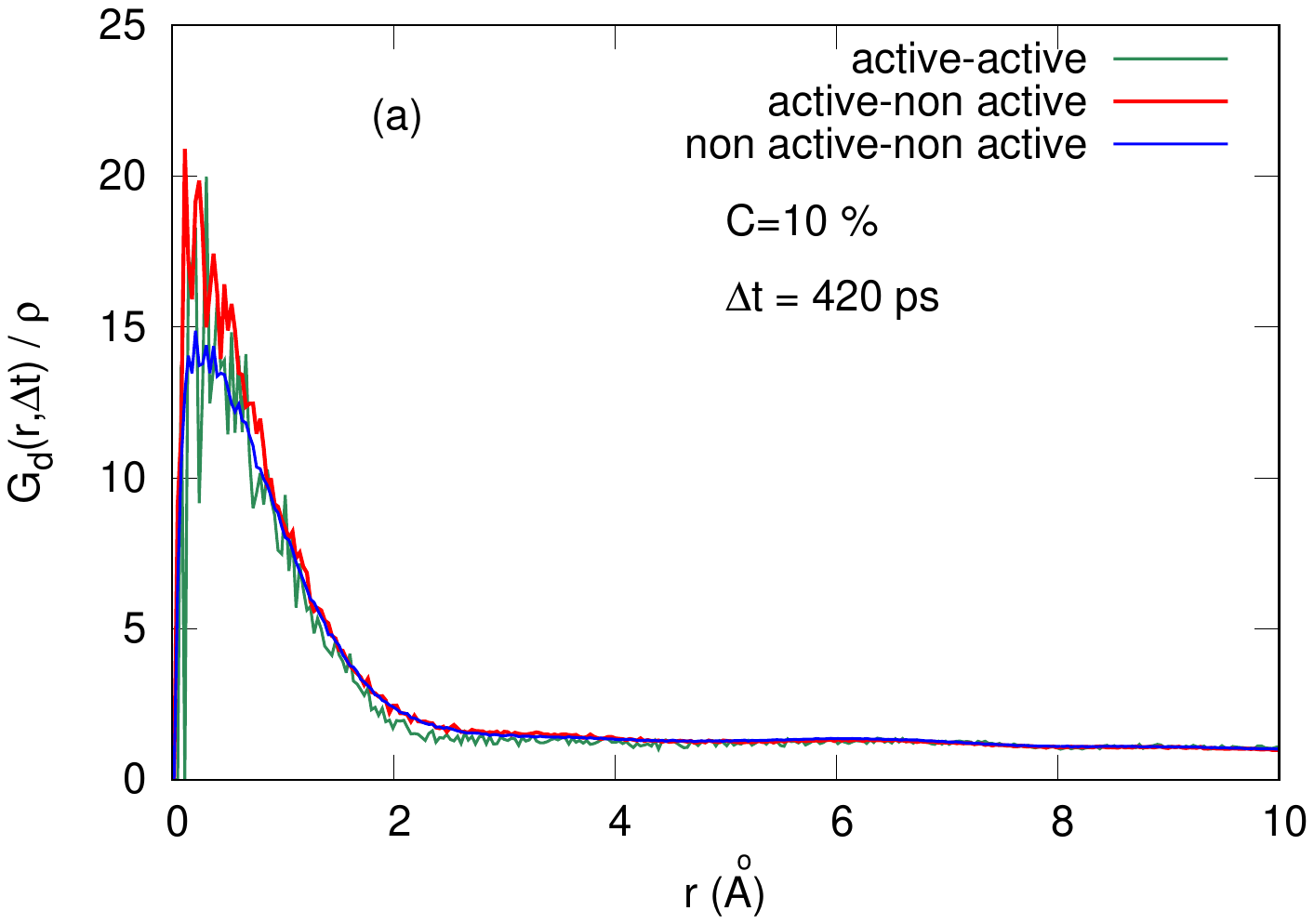}
\includegraphics[height=7. cm]{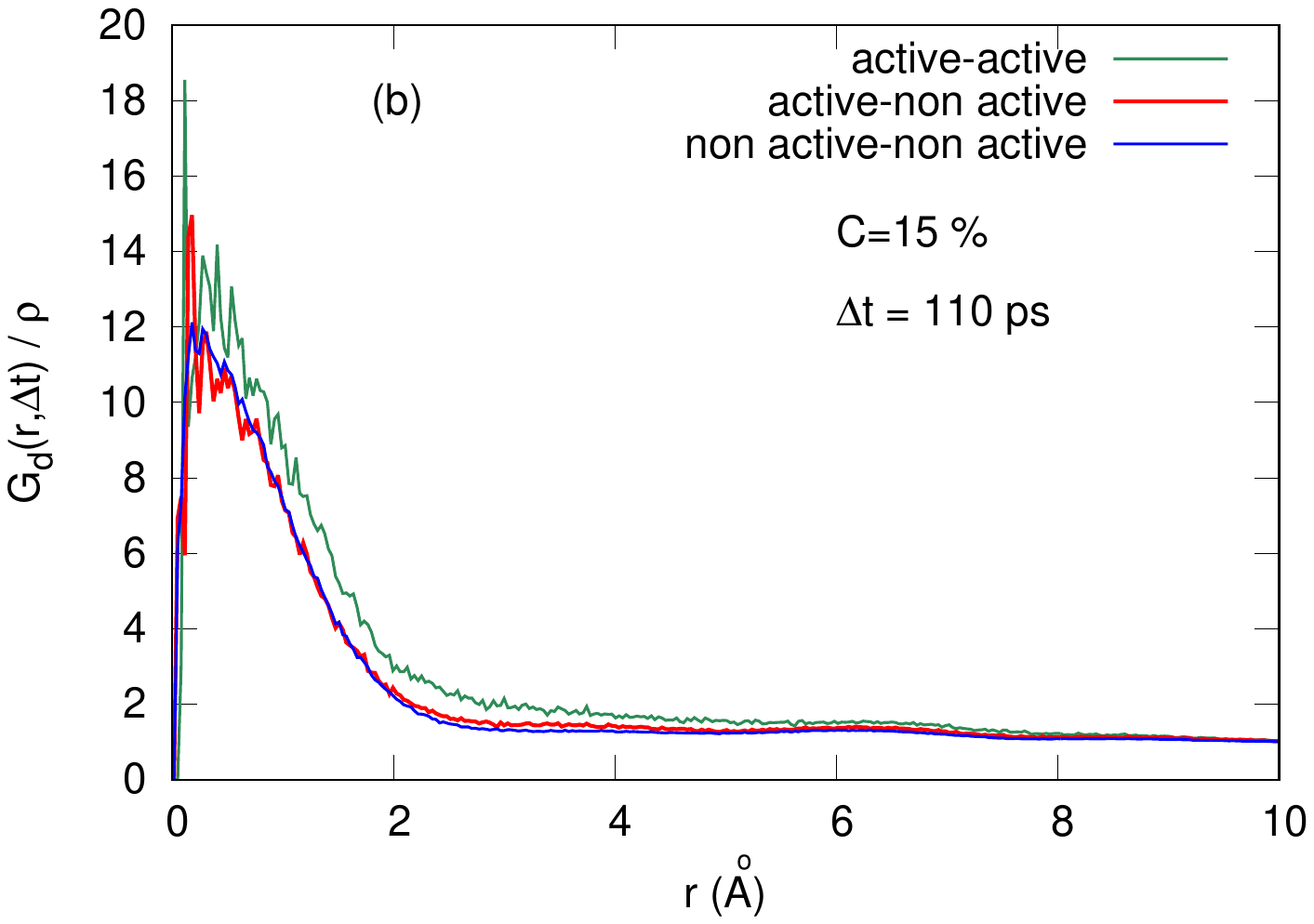}
\includegraphics[height=7. cm]{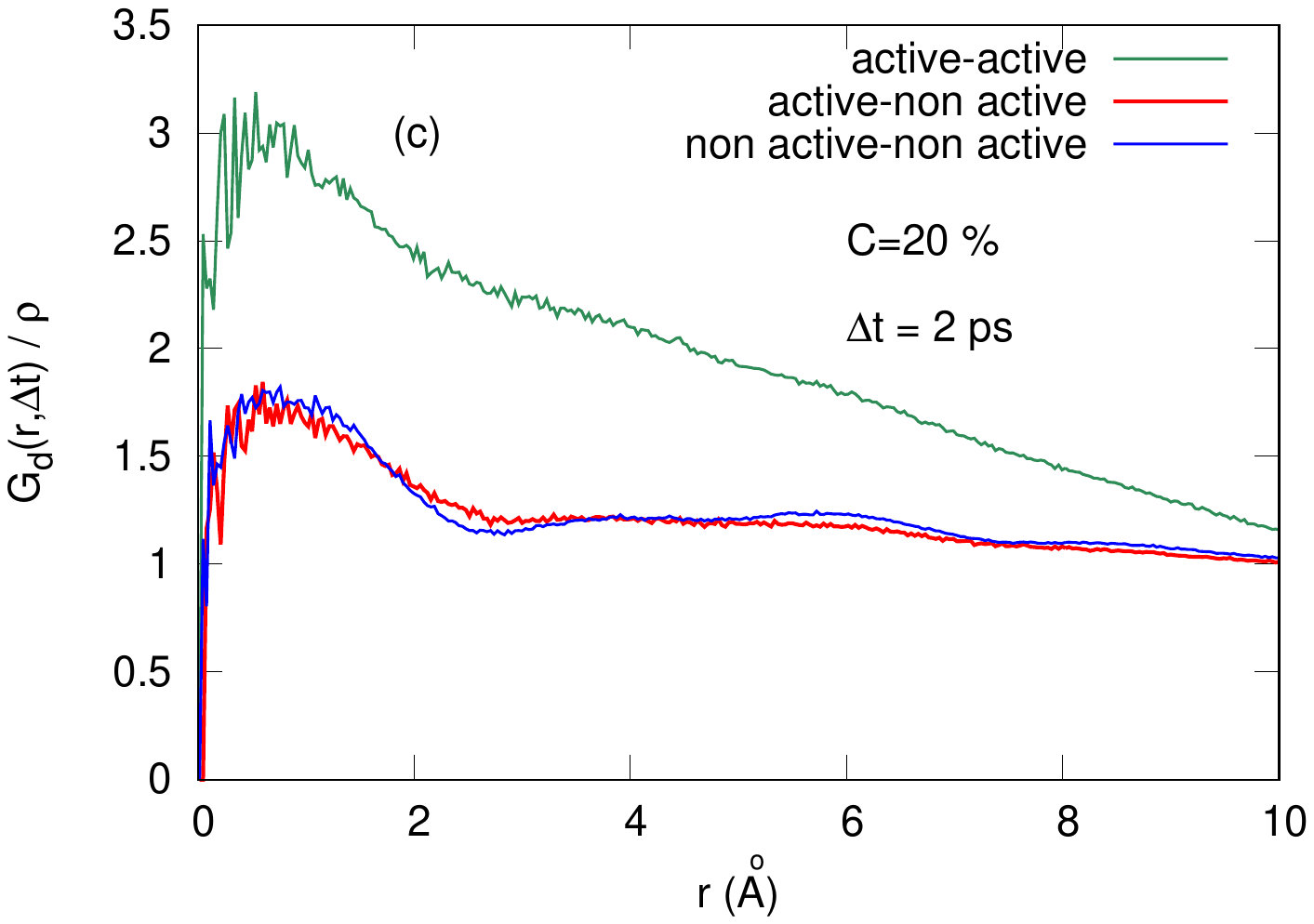}

\caption{(color online)  Distinct part of the Van Hove correlation function for the $6 \%$ most mobile molecules. (a) below the transition, (b) around the transition, (c) above the transition. 
We observe a peak around $r=0$. This peak is a characteristic of string like cooperative motions of the particles, showing that a molecule replaces the former on the characteristic time. The large peak in the Van Hove between active and non-active molecules in (a) shows non active molecules replacing active ones on a  characteristic time $\Delta t=450ps$. The motion of active molecules is thus transmitted to non active ones via the cooperative mechanism of the medium. }
\label{f8a}
\end{figure}

\section{Conclusion}

{\color{black} Due to the cage effect, in supercooled liquids the velocity is not a relevant physical parameter as a molecule bounces several times on the cage boundary, losing its initial direction  before moving outside the cage. For this reason and because the mobility has been established as a control parameter for spontaneous cooperative motions, we used the mobility instead of the velocity in our model of active matter. We found a transition for a concentration of active molecules that decreases significantly when the time parameter of the mobility is large enough. As the instantaneous velocity is equivalent to a mobility with a time parameter null, the observed transition will not appear if the velocity was used instead of the mobility.}

We found that the mobility of active molecules is transmitted to inactive ones via the medium's cooperative mechanisms. This result is in agreement with facilitation theories.
{\color{black} There is a facilitation of motion around mobile molecules because the motion of a molecule generates a void thus opening a cage that permits the motion of nearby molecules. The facilitation induces an aggregation of mobile molecules and a fluidization around them.
The transmission then continues from inactive ones to other inactive molecules via the cooperative mechanisms until the transition is reached if the threshold of active molecules concentration is attained. That is when the concentration of active molecules is large enough to induce the mobility of most molecules.}

{\color{black} The decrease of the excitation concentration when the temperature drops has been postulated as the origin of the increase of cooperative mechanisms in supercooled liquids. However, in opposition with that view, we found that the cooperative mechanisms do not decrease when the concentration of active molecules, then the concentration of excitations increases, as long as the temperature is maintained constant.}
This result shows that the cooperative mechanisms are not induced or directly related to the decrease of the excitation concentration in supercooled liquids when the temperature decreases.

Spontaneous cooperative motions are more probably induced by the decrease of the kinetic energy (temperature) that induces a preponderance of the potential energy, therefore an organization of the medium. 
{\color{black} As the temperature decreases,  kinetic energy per molecule becomes no longer enough to break the cage and enable the molecule's motion. For the motion to still occur, neighbor molecules have to move simultaneously. The kinetic energy and also the potential energy that defines the strength of the caging appear therefore in that view as the significant parameters for cooperative motions.

Finally, we found that when $\tau_{m}$ is large enough, the critical concentration that trigger the fluidization transition is small. The transition could then be used in various applications, as thermophoresis, nano-structuration of materials or in petroleum extraction processes.}

 }

\vskip 0.5cm

{\bf \Large  Conflict of interest}

There are no conflict of interest to declare.

\vskip 0.5cm

{\bf \Large  Data availability}

The data that support the findings of that study are available from the corresponding author upon reasonable request.

\end{document}